\documentclass[preprint,preprintnumbers, prd, floatfix, superscriptaddress,nofootinbib] {revtex4-1}
\usepackage{epsfig}
\usepackage{subfigure}
\usepackage{dcolumn}
\usepackage{bm}
\usepackage[usenames ,dvipsnames]{xcolor}
\usepackage{slashed}
\usepackage{graphicx,color}
\usepackage{hyperref}

\begin{document}
\title{Rescattering-induced $D\to SS$ weak decays}

\author{Yan-Li Wang}
\email{ylwang0726@163.com}
\affiliation{School of Physics and Electronic Engineering, Shanxi Normal University,
Taiyuan 030031, China}

\author{Shu-Ting Cai}
\email{18734581917@163.com}
\affiliation{School of Physics and Electronic Engineering,
Shanxi Normal University, Taiyuan 030031, China}

\author{Yu-Kuo Hsiao}
\email{yukuohsiao@gmail.com}
\affiliation{School of Physics and Electronic Engineering,
Shanxi Normal University, Taiyuan 030031, China}

\date{\today}

\begin{abstract}
We investigate two-body non-leptonic $D\to SS$ weak decays,
where $S$ denotes a light scalar meson such as $a_0/a_0(980)$, $f_0/f_0(980)$,
or $\sigma_0/f_0(500)$. 
Short-distance topologies from $W$-boson emission and annihilation (exchange) 
are found to be negligible, while long-distance final-state interactions 
provide the dominant contributions.
In particular, triangle rescattering processes, $D \to \pi\eta^{(\prime)} \to \sigma_0 a_0$ 
and $D \to a_1(1260)\eta \to \sigma_0 a_0$, mediated by pion exchange 
in $\pi\eta^{(\prime)}$ and $a_1(1260)\eta$ scatterings, respectively, 
are identified as the leading mechanisms. Our calculations yield branching fractions 
${\cal B}(D_s^+ \to \sigma_0 a_0^+) = (1.0 \pm 0.2^{+0.1}_{-0.2}) \times 10^{-2}$, 
${\cal B}(D^+ \to \sigma_0 a_0^+) = (1.1 \pm 0.2^{+0.1}_{-0.2}) \times 10^{-3}$, and 
${\cal B}(D^0 \to \sigma_0 a_0^0) = (0.9 \pm 0.2^{+0.2}_{-0.3}) \times 10^{-5}$. 
For the Cabibbo-allowed decay mode $D_s^+ \to f_0 a_0^+$, 
the near-threshold condition $m_{D_s}\simeq m_{f_0}+m_{a_0}$ limits the phase space, 
suppressing the branching fraction to $(3.4\pm0.3^{+0.4}_{-0.9})\times 10^{-4}$.
These results highlight rescattering-induced $D\to SS$ decays 
as promising channels for experimental studies at BESIII, Belle(-II), and LHCb.
\end{abstract}

\maketitle
\section{introduction}
Two-body non-leptonic $D\to MM$ decays, 
where the final-state meson $M$ can be either a psedoscalar $(P)$ or a vector $(V)$,
have long attracted considerable interest in both theoretical and experimental studies.
These decays provide valuable insights into hadronization dynamics of weak interactions
and serve as an important testing ground for studies of $CP$ violation~\cite{Bhattacharya:2009ps,Cheng:2010ry,Bhattacharya:2010uy,
Fusheng:2011tw,Bhattacharya:2012ah,Cheng:2012wr,Brod:2012ud,Li:2013xsa,
Cheng:2012xb,Cheng:2019ggx,Cheng:2024hdo}. 
The final-state meson may also be a light scalar meson ($S$), 
such as $a_0/a_0(980)$, $f_0/f_0(980)$, $\sigma_0/f_0(500)$, or $\kappa/K_0^*(700)$,
which are conventionally regarded as p-wave $q\bar q$ states~\cite{Anisovich:2000wb,Bediaga:2003zh,Aliev:2007uu,Colangelo:2010bg,
Shi:2015kha,Soni:2020sgn,Klempt:2021nuf}.
At the same time, alternative non-$q\bar q$ interpretations have been widely discussed, 
in which the light scalar mesons are described either as compact $q^2\bar q^2$ states~\cite{Jaffe:1976ig,
Jaffe:1976ih,Close:2002zu,Pelaez:2003dy,Maiani:2004uc,Amsler:2004ps,
Jaffe:2004ph,Achasov:2005hm,tHooft:2008rus,Fariborz:2009cq,
Weinberg:2013cfa,Agaev:2018fvz,Hsiao:2023qtk,Hsiao:2025sfz,Wang:2025uie} or as
meson–meson molecular configurations~\cite{Weinstein:1990gu,Branz:2007xp,Baru:2003qq,
Dai:2012kf,Dai:2014lza,Sekihara:2014qxa,Yao:2020bxx,Wang:2022vga}. In this context, the decays
$D\to PS$ and $D\to VS$ have proven to be particularly valuable 
in helping to resolve the longstanding debate concerning the quark composition 
of light scalar mesons~\cite{Cheng:2002ai,Cheng:2022vbw,Yu:2021euw,
Hsiao:2019ait,Hsiao:2019ait,Ling:2021qzl,Achasov:2024nrh,Cheng:2024zul}.

In principle, $D\to SS$ decays can also serve as a probe of the internal structure 
of light scalar mesons. However, the short-distance $W$-emission topologies that
typically dominate $D\to MM$ decays are strongly suppressed in these channels 
due to the vanishing or nearly zero scalar decay constants 
$f_S$~\cite{Cheng:2002ai,Cheng:2022vbw}. 
These decay constants govern scalar meson production from the vacuum
through the matrix elements $\langle S|\bar q_1\gamma_\mu(1-\gamma_5) q_2|0\rangle=f_S q_\mu$.
As a consequence, the corresponding branching fractions 
are expected to be small, suggesting that experimental studies may be challenging.

Certain $D\to SP$ and $D\to SV$ decays,
such as $D_s^+\to a_0^{+(0)}\pi^{0(+)}$ and $D_s^+\to a_0^{+}\rho^0$~\cite{Hsiao:2019ait,Yu:2021euw},
are likewise expected to exhibit small branching fractions as a result of the suppression of
both $W$-emission and $W$-exchange topologies~\cite{Hsiao:2014zza,Hsiao:2019wyd,Hsiao:2023mud}.
Nevertheless, the measured branching fractions, $(2.1 \pm 0.4)\times 10^{-2}$~\cite{pdg,Ablikim:2019pit}
and $(2.1 \pm 0.9)\times 10^{-3}$~\cite{pdg,BESIII:2021aza}, respectively, 
are significantly larger than short-distance expectations. 
To account for this discrepancy, long-distance effects, most notably final-state interactions~(FSI)~\cite{Fajfer:2003ag,Hsiao:2019ait,Yu:2021euw,Yu:2020vlt,
Ling:2021qzl,Hsiao:2021tyq,Hsiao:2024szt}, have been proposed. In particular, 
triangle rescattering processes have been shown to significantly enhance the decay amplitudes~\cite{Hsiao:2019ait,Yu:2021euw}, 
thereby reconciling the discrepancy between theoretical expectations  and experimental observations. 
Similar long-distance enhancement mechanisms may also play an important role in $D \to SS$ decays, 
which, however, remain largely unexplored.

To illustrate the FSI mechanism, consider the decay
$D_s^+\to \pi^+ \eta$, where the final-state mesons
$\pi^+$ and $\eta$ can rescatter via the charged pion exchange,
converting into $\rho^0$ and $a_0^+$, respectively.
This triangle rescattering mechanism significantly enhances the branching fraction
${\cal B}(D_s^+\to \rho^0 a_0^+)$ to the $10^{-3}$ level~\cite{Yu:2021euw}.
By replacing the intermediate $\rho^0 \to \pi^+\pi^-$ transition with $\sigma_0\to\pi^+\pi^-$
at one of the triangle vertices, a similar process can generate the scalar–scalar final state
$\sigma_0 a_0^+$. Consequently, the decay $D_s^+\to\sigma_0 a_0^+$
emerges as a promising channel, closely analogous to the observed $D_s^+\to \rho^0 a_0^+$.

In this work, we investigate FSI rescattering-induced decays $D\to \sigma_0 a_0$,
where the parent meson can be $D_s^+$, $D^+$, and $D^0$.
We further examine the potential decay $D_s^+\to  f_0 a_0^+$.
Our study provides the first predictions for $D\to SS$ decay channels and 
underscores the crucial role of long-distance FSI mechanisms in shaping their decay dynamics.

\section{Formalism}
%
\begin{figure}[t!]
\includegraphics[width=2.4in]{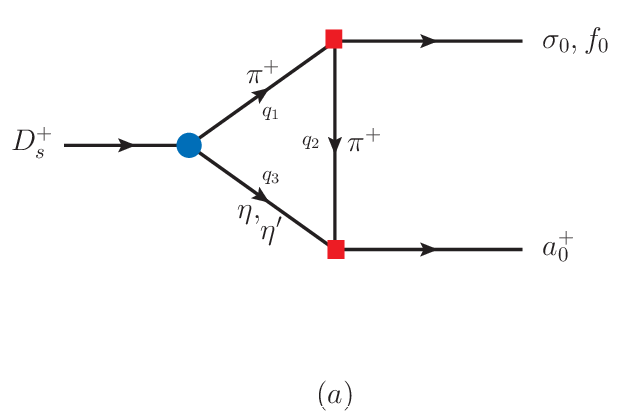}
\includegraphics[width=2.4in]{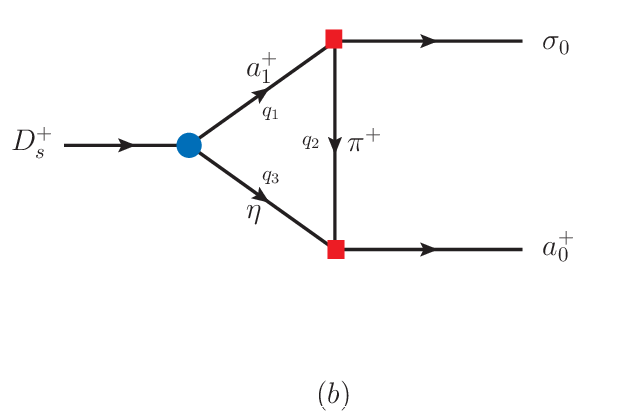}
\includegraphics[width=2.4in]{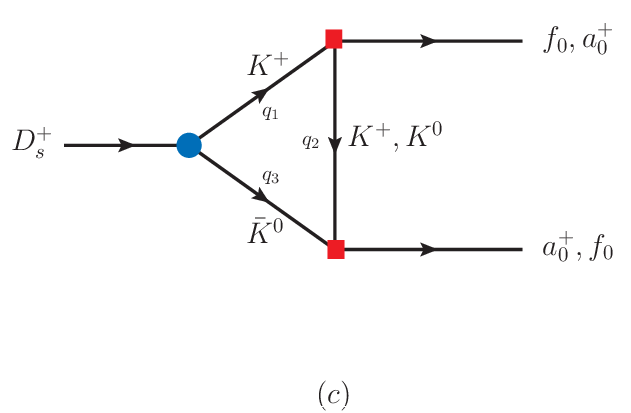}
\caption{Triangle rescattering diagrams are considered, 
where (a,b) correspond to $D_s^+ \to (\pi^+\eta^{(\prime)}, a_1^+\eta) \to \sigma_0 a_0^+$ 
mediated by charged pion exchange, while (a,c) represent $D_s^+ \to (\pi^+\eta^{(\prime)}, 
K^+\bar K^0) \to f_0 a_0^+$ mediated by $\pi^+$ and $K^+(\bar K^0)$ exchanges, 
respectively.}\label{fig1}
\end{figure}
\begin{figure}[t!]
\includegraphics[width=2.4in]{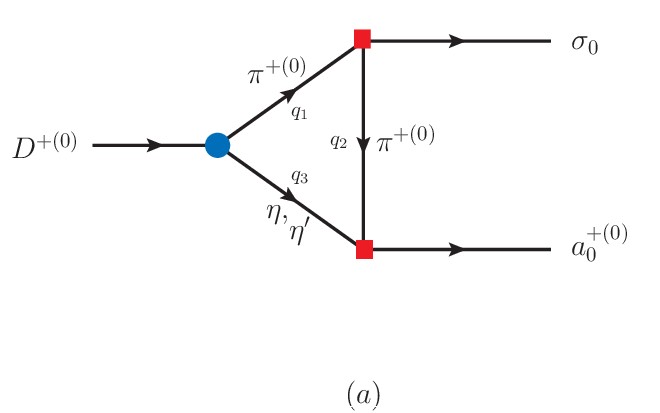}
\includegraphics[width=2.4in]{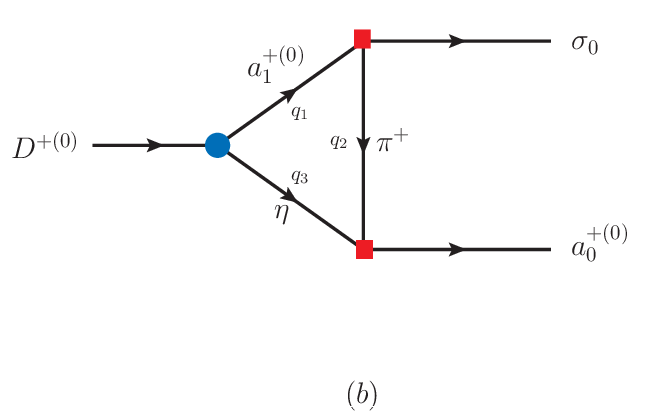}
\caption{Triangle rescattering diagrams, where (a) and (b) illustrate the decays
$D^{+(0)}\to (\pi^{+(0)}\eta,\pi^{+(0)}\eta^\prime)\to\sigma_0 a_0^{+(0)}$ and
$D^{+(0)}\to a_1^{+(0)}\eta\to \sigma_0 a_0^{+(0)}$, respectively,
mediated by pion exchange in the triangle loops.}\label{fig2}
\end{figure}

%
The long-distance triangle rescattering mechanism responsible for scalar–scalar
$(SS)$ pair formation originates from tree-level two-body $D\to PP$ weak decays.
For $D_s^+\to\sigma_0 a_0^+$, the $D_s^+$ meson first decays into
$\pi\eta^{(\prime)}$ and $a_1\eta$, as illustrated in Figs.~\ref{fig1}$(a)$~and~\ref{fig1}$(b)$,
respectively. These intermediate states then rescatter via charge pion exchange,
converting into the $SS$ pair $\sigma_0 a_0^+$. In a similar manner,
Figs.~\ref{fig1}(a,c) depict the rescattering pathway 
$D_s^+ \to (\pi \eta^{(\prime)}, K^+ \bar K^0) \to f_0 a_0^+$, 
while Figs.~\ref{fig2}(a,b) show the corresponding mechanisms 
for $D^{+(0)} \to (\pi^{+(0)} \eta^{(\prime)}, a_1^{+(0)} \eta) \to \sigma_0 a_0^{+(0)}$.
The amplitudes of the initial weak decays $D \to \pi^+ \eta,\pi^+ \eta^{\prime}$
and $D_s^+\to \bar K^0 K^+$ are
given by~\cite{Cheng:2010ry,Cheng:2024hdo,Cheng:2019ggx}
\begin{eqnarray}\label{amp1}
{\cal M}(D_s^+\to \pi^+\eta)&=&\lambda_{sd}(\sqrt 2 A\cos\phi-T\sin\phi)\,,
\nonumber\\
{\cal M}(D_s^+\to \pi^+\eta')&=&\lambda_{sd}(\sqrt 2 A\sin\phi+T\cos\phi)\,,
\nonumber\\
{\cal M}(D_s^+\to \bar K^0 K^+)&=&\lambda_{sd}(A+C)\,,
\nonumber\\
{\cal M}(D^+\to \pi^+\eta)&=&
\frac{1}{\sqrt 2}\lambda_d (T_d+C_d+A_{\eta})\cos\phi-\lambda_s C_s\sin\phi\,,
\nonumber\\
{\cal M}(D^+\to \pi^+\eta')&=&
\frac{1}{\sqrt 2}\lambda_d (T_d+C_d+A_{\eta'})\sin\phi+\lambda_s C_s\cos\phi\,,
\nonumber\\
{\cal M}(D^0\to \pi^0\eta)&=&-\lambda_d E_d\cos\phi-\frac{1}{\sqrt 2}\lambda_s C_s\sin\phi\,,
\nonumber\\
{\cal M}(D^0\to \pi^0\eta')&=&-\lambda_d E_d\sin\phi+\frac{1}{\sqrt 2}\lambda_s C_s\cos\phi\,,
\end{eqnarray}
where $\lambda_{sd}\equiv V_{cs}^* V_{ud}$, $\lambda_d\equiv V_{cd}^* V_{ud}$,
and $\lambda_s\equiv V_{cs}^* V_{us}$, with $V_{ij}$
denoting the Cabibbo-Maskawa-Kabayashi (CKM) matrix elements.
In the amplitudes, $\sin\phi$ and $\cos\phi$ account for the $\eta-\eta'$ mixing~\cite{FKS,FKS2}, 
where 
$\eta=\cos\phi |\sqrt{1/2}(u\bar u+d\bar d)\rangle-\sin\phi |s\bar s\rangle$ and
$\eta'=\sin\phi |\sqrt{1/2}(u\bar u+d\bar d)\rangle+\cos\phi |s\bar s\rangle$,
with the mixing angle $\phi=40.4^\circ$~\cite{Cheng:2024hdo}.
The parameters $T_{(q)}$, $C_{(q)}$, $A$, $A_{\eta^{(\prime)}}$
and $E_{(q)}$ (with $q=d,s$) represent the topological amplitudes
based on $SU(3)$ flavor $[SU(3)_f]$ symmetry, with the Fermi constant
absorbed into their definitions.

%
%
From ${\cal B}(D_s^+\to a_1^+\eta,a_1^+\to\sigma_0\pi^+,\sigma_0\to\pi^+\pi^-)
=(2.5\pm 0.9)\times 10^{-3}$ with the axial-vector meson $a_1\equiv a_1(1260)$~\cite{pdg}, 
together with
${\cal B}(a_1^+\to\sigma_0\pi^+)\simeq (18.76\pm 4.29\pm 1.48)\times 10^{-2}$~\cite{pdg} and
${\cal B}(\sigma_0\to \pi^+\pi^-)=0.67$~\cite{Cheng:2002ai,Cheng:2022vbw,Wang:2025uie},
we extract ${\cal B}(D_s^+\to a_1^+\eta)=(2.0\pm 0.9)\times 10^{-2}$.
This branching fraction at the $10^{-2}$ level suggests that
$D_s^+ \to a_1^+ \eta$ can serve as a significant source for $D_s^+ \to \sigma_0 a_0^+$, 
with the strong decay $a_1 \to \sigma_0 \pi$ naturally entering the triangle loop [see Fig.~\ref{fig1}(b)]. 
Analogous rescattering processes, such as those depicted in Fig.~\ref{fig2}(b), 
can also contribute to $D^{+(0)} \to \sigma_0 a_0^{+(0)}$.
The corresponding amplitudes are~\cite{Cheng:2007mx}
\begin{eqnarray}\label{amp2}
\hat{\cal M}(D_s^+\to a_1^+\eta)&=&
\lambda_{sd} a_{1{\rm w}} f_{a_1} F_1^{D_s^+\to\eta}(m_{a_1}^2)\,,\nonumber\\
\hat{\cal M}(D^+\to a_1^+\eta)&=&
\lambda_d a_{1{\rm w}} f_{a_1} F_1^{D^+\to\eta}(m_{a_1}^2)
+(\lambda_d f_{\eta}^d/\sqrt 2+\lambda_s f_{\eta}^s)
a_{2{\rm w}} V_0^{D^+\to a_1^+}(m_{\eta}^2)\,,\nonumber\\
\hat{\cal M}(D^0\to a_1^0\eta)&=&
(\lambda_d f_{\eta}^d/\sqrt 2+\lambda_s f_{\eta}^s)a_{2{\rm w}} V_0^{D^0\to a_1^0}(m_{\eta}^2)\,,
\end{eqnarray}
where ${\cal M}\equiv (G_F/\sqrt 2)(2 m_{a_1}\varepsilon\cdot p_{D_{(s)}})\hat{\cal M}$, and
$\varepsilon_\mu$ is the polarization four-vector of the axial-vector meson.
The factorization parameter $a_{1{\rm w}(2{\rm w})} = c_{1(2)} + c_{2(1)}/N_c$ 
combines the Wilson coefficients $c_{1,2}$ and the color number $N_c$.
The transition form factors $F_1^{D\to\eta}(q^2)$ and $V_0^{D\to a_1}(q^2)$ are taken from Refs.~\cite{Soni:2018adu,Cheng:2003sm}, with isospin symmetry implying 
$V_0^{D^+\to a_1^+} = \sqrt{2}\,V_0^{D^0\to a_1^0}$.
The decay constants $f_{\eta}^{d}$ and $f_{\eta}^{s}$ are defined by
$\langle \eta|\bar d\gamma_\mu \gamma_5 d|0\rangle=-i(f^d_{\eta}/\sqrt 2) q_\mu$
and $\langle \eta|\bar s\gamma_\mu \gamma_5 s|0\rangle
=-if^s_{\eta} q_\mu$~\cite{Beneke:2002jn}, respectively.
For the strong decays $f_0\to\pi\pi(K\bar K)$, $a_0\to\pi\eta(K\bar K)$, and $a_1\to \sigma_0\pi$,
the amplitudes are~\cite{Hsiao:2025sfz,Hsiao:2023qtk}
\begin{eqnarray}\label{amp3}
&&
{\cal M}(\sigma_0\to\pi^+\pi^-)=\sqrt 2 {\cal M}(\sigma_0\to\pi^0\pi^0)
=g_{\sigma_0\pi\pi}\,,\nonumber\\
&&
{\cal M}(f_0\to\pi^+\pi^-)=\sqrt 2 {\cal M}(f_0\to\pi^0\pi^0)
=g_{f_0\pi\pi}\,,
\nonumber\\
&&
{\cal M}(a_0^{+(0)}\to\pi^{+(0)}\eta)=g_{a_0\pi\eta}\,,
\nonumber\\
&&
{\cal M}(a_1^{+(0)}\to\sigma_0\pi^{+(0)})=g_{a_1\sigma_0 \pi}\varepsilon\cdot(p_{\sigma_0}-p_\pi)\,,
\nonumber\\
&&
{\cal M}(f_0\to K^+ K^-(K^0\bar K^0))=g_{f_0 K\bar K}\,,
\nonumber\\
&&
{\cal M}(a_0^+\to K^+\bar K^0)=(-)\sqrt 2{\cal M}(a_0^0\to K^+ K^-(K^0\bar K^0))
=g_{a_0 K\bar K}\,,
\end{eqnarray}
where $g_{SPP}$ and $g_{APP}$ denote the relevant strong coupling constants.
In particular, the $SU(3)_f$ relations imply the quark compositions
$(a_0^+,a_0^0)=(|u\bar d \rangle,|\sqrt {1/2}(u\bar u-d\bar d)\rangle)$ in the $q\bar q$ configuration,
and $(a_0^+,a_0^0)=(|u\bar d s\bar s\rangle,|\sqrt {1/2}(u\bar u-d\bar d)s\bar s\rangle)$
in the $q^2\bar q^2$ configuration~\cite{Stone:2013eaa,Maiani:2004uc}.
As part of the $a_0^0$ quark content, 
the component $-\sqrt {1/2}d\bar d$ or $-\sqrt {1/2}d\bar d s\bar s$ 
give rise to a compensating factor of $-\sqrt 2$ in the amplitude for 
$a_0^0\to K^0(\bar s d)\bar K^0(s\bar d)$, as presented in Eq.~(\ref{amp3}).
It is worth noting that, in the $q\bar q$ configuration,
the $s\bar s$ pair appearing in the $K^0\bar K^0$ final state 
constitutes an additional component that must be generated via vacuum-induced gluon splitting, 
$g\to s\bar s$. In contrast, in the $q^2\bar q^2$ configuration, the required $s\bar s$
component is already present intrinsically and can be naturally rearranged into
the $K^0\bar K^0$ final state.

Using Eqs.~(\ref{amp1}), (\ref{amp2}), and (\ref{amp3}),
the triangle rescattering amplitudes corresponding to Fig.~\ref{fig1}(a,b,c)
are expressed as~\cite{Hsiao:2019ait,Yu:2021euw}
\begin{eqnarray}\label{amp_abc}
&&
{\cal M}^{(\prime)}_{a(\sigma_0,f_0)}=\int \frac{d^4{q}_{1}}{(2\pi)^{4}}
\frac{{\cal M}_{D_s^+\to\pi^+\eta^{(\prime)}}{\cal M}_{(\sigma_0,f_0)\to\pi^+\pi^-}
{\cal M}_{a_0^+\to\pi^+\eta^{(\prime)}}F_\pi(q_{2}^2)}
{(q_{1}^{2}-{m_1}^{2})[(q_1-p_1)^{2}-m_{2}^{2}][(q_1-p_0)^{2}-m_{3}^{2}]}\,,\nonumber\\
&&
{\cal M}_b=\int \frac{d^4{q}_{1}}{(2\pi)^{4}}
\frac{{\cal M}_{D_s^+\to a_1^+\eta}{\cal M}_{a_1^+\to\sigma_0\pi^+}
{\cal M}_{a_0^+\to\pi^+\eta}F_\pi(q_{2}^2)}
{(q_{1}^{2}-{m_1}^{2})[(q_1-p_1)^{2}-m_{2}^{2}][(q_1-p_0)^{2}-m_{3}^{2}]}\,,\nonumber\\
&&
{\cal M}_{c}=2\int \frac{d^4{q}_{1}}{(2\pi)^{4}}
\frac{{\cal M}_{D_s^+\to K^+\bar K^0}{\cal M}_{f_0\to K^+ K^-}
{\cal M}_{a_0^+\to K^+\bar K^0}F_K(q_{2}^2)}
{(q_{1}^{2}-{m_1}^{2})[(q_1-p_1)^{2}-m_{2}^{2}][(q_1-p_0)^{2}-m_{3}^{2}]}\,,
\end{eqnarray}
where ${\cal M}_{A\to BC}\equiv {\cal M}(A\to BC)$, and we define
${\cal M}^{(\prime)}_{a(\sigma_0,f_0)}={\cal M}(D_s^+\to\pi^+\eta^{(\prime)}\to\sigma_0 a_0^+,f_0 a_0^+)$,
${\cal M}_b={\cal M}(D_s^+\to a_1^+\eta\to \sigma_0 a_0^+)$, and
${\cal M}_c={\cal M}(D_s^+\to K^+\bar K^0\to f_0 a_0^+)$.
The factor of 2 in ${\cal M}_c$ accounts for the contributions of the two different exchanged kaons 
($K^+$ and $K^0$).  Here, $q_1$, $q_2$, and $q_3$ denote the internal loop momenta 
assigned in Fig.~\ref{fig1}(a,b,c), while $p_0^\mu$ and $(p_1^\mu,p_2^\mu)$ 
are the four-momenta of the initial and final states, respectively.
The form factors $F_{\pi,K}(q_{2}^2)\equiv
(\Lambda_{\pi,K}^{2}-m^{2}_2)/(\Lambda_{\pi,K}^{2}-q^{2}_{2})$
regularize the triangle-loop integrals~\cite{Du:2021zdg}.
The total rescattering amplitudes are then
\begin{eqnarray}
&&
{\cal M}(D_s^+\to \sigma_0 a_0^+)={\cal M}_{a(\sigma_0)}+{\cal M}'_{a(\sigma_0)}+{\cal M}_b\,,\nonumber\\
&&
{\cal M}(D_s^+\to f_0 a_0^+)={\cal M}_{a(f_0)}+{\cal M}'_{a(f_0)}+{\cal M}_c\,,
\end{eqnarray}
where ${\cal M}'_{a(\sigma_0,f_0)}={\cal M}(D_s^+\to\pi^+\eta'\to\sigma_0 a_0^+,f_0 a_0^+)$
accounts for the $\eta'$ contribution in Fig.~\ref{fig1}(a).
Analogously, Figs.~\ref{fig2}(a,b) lead to the total amplitude 
${\cal M}(D^{+(0)} \to \sigma_0 a_0^{+(0)})$.

\section{Numerical analysis}
In our numerical analysis, we adopt the Wolfenstein parameterization for 
the CKM matrix elements
$(V_{cs}, V_{cd})=(1-\lambda^2/2,-\lambda)$ and
$(V_{ud}, V_{us})=(1-\lambda^2/2,\lambda)$
with $\lambda=0.22453\pm 0.00044$~\cite{pdg}.
The extracted topological parameters in Eq.~(\ref{amp1})
are taken from Ref.~\cite{Cheng:2024hdo},
given by
\begin{eqnarray}\label{TCA}
&&
(T, C)
=[3.13\pm 0.01,(2.58\pm 0.01)e^{-i(151.9\pm 0.3)^{\circ}}]\times 10^{-6}\,\mbox{GeV}\,,\nonumber\\
&&
(E,A)
=[(1.47\pm 0.02)e^{i(121.7\pm 0.4)^{\circ}},
(0.39\pm 0.02)e^{i(14.1^{+11.0}_{-\,\;8.5})^{\circ}}]\times 10^{-6}\,\mbox{GeV}\,.
\end{eqnarray}
To account for $SU(3)_f$ symmetry breaking~\cite{Cheng:2019ggx,Cheng:2012xb},
we adopt the relations
$(T_d,T_s)=(0.82,1.27)\times T$,
$(C_d,C_s)=(0.93,1.29)\times C$,
$(E_d,E_s)=(1.24 e^{i 13.7^\circ},1.55 e^{-i 12.3^\circ})\times E$,
and $(A_\eta,A_{\eta'})=(1.15,1.56)\times A$.

The $D\to a_1\eta$ decays involve the decay constants
$|f_{a_1}|=(0.203\pm 0.018)$~GeV~\cite{Cheng:2003sm,Bloch:1999vka},
$(f^d_\eta,f^s_\eta)=(0.108,-0.111)$~GeV~\cite{Hsiao:2018umx,Beneke:2002jn},
and the factorization parameters
$(a_{1{\rm w}},a_{2{\rm w}})=(0.93\pm 0.04, 0.37\pm 0.10)$~\cite{Hsiao:2019ait}.
The form factors are represented as~\cite{Soni:2018adu,Cheng:2003sm}
\begin{eqnarray}
&&
F_1^{D_{(s)}^+\to\eta}(q^2)=
\frac{F_1^{D_{(s)}^+\to\eta}(0)}{1-a(q^2/m_{D_{(s)}}^2)+b(q^4/m_{D_{(s)}}^4)}\,,
\nonumber\\
&&
V_0^{D^+\to a_1^+}(q^2)=
\frac{V_0^{D^+\to a_1^+}(0)}{1-a(q^2/m_{D}^2)+b(q^4/m_{D}^4)}\,,
\end{eqnarray}
with
$(F_1^{D_{s}^+\to\eta}(0),\,a,\,b)=(0.78,\,0.69,\,0.002)$,
$(F_1^{D^+\to\eta}(0),\,a,\,b)=(0.67,\,0.93,\,0.12)$, and
$(V_0^{D^+\to a_1^+}(0),\,a,\,b)=(0.31,\,0.85,\,0.49)$.

The strong coupling constants
are given by~\cite{Bugg:2008ig,Kornicer:2016axs,pdg,BESIII:2020hfw}
\begin{eqnarray}\label{couplings}
&&
g_{\sigma_0\pi\pi}=(3.27\pm 0.05)~\text{GeV}\,,\;g_{a_1\sigma_0 \pi}=4.2\pm 0.5\,,\;\nonumber\\
&&
(g_{f_0\pi\pi},g_{f_0 K\bar K})=(1.5\pm 0.1,3.54\pm 0.05)~\text{GeV}\,,\nonumber\\
&&
(g_{a_0\pi\eta},g_{a_0\pi\eta'},g_{a_0 K\bar K})=(2.87\pm 0.09,-2.52\pm 0.08,2.94\pm 0.13)~\text{GeV}\,.
\end{eqnarray}
The couplings $g_{\sigma_0\pi\pi}$ and $g_{a_1\sigma_0 \pi}$ are extracted from
Eq.~(\ref{amp3}) and the available data in~\cite{pdg}, with
$g_{\sigma_0\pi\pi}=[(8\pi m_{\sigma_0}^2/|\vec{p}_{\rm cm}|)(2/3)\Gamma_{\sigma_0}^0]^{1/2}$ and $g_{a_1\sigma_0 \pi}=(6\pi m_{a_1}^2 \Gamma_{a_1\to \sigma_0\pi}/|\vec{p}_{\rm cm}|^3)^{1/2}$.

In Eq.~(\ref{amp_abc}), the cutoff parameter
$\Lambda_{M}$ is a phenomenological input whose value is inferred from experimental data.
According to Refs.~\cite{Tornqvist:1993ng,Li:1996yn,Wu:2019vbk,Yu:2021euw},
$\Lambda_{M}$ is typically of order 1.0~GeV.
An additional relation, $\Lambda_K-\Lambda_\pi=m_K-m_\pi$,
can be imposed as a constraint in our analysis~\cite{Cheng:2004ru}. Furthermore,
the measured branching fractions of 
$(D^+,D_s^+,D^0)\to \pi a_0$~\cite{BESIII:2019jjr,BESIII:2024tpv},
which comprise six independent data points, 
provide further guidance for determining more precise cutoff values.
Since these decays proceed via $\rho\eta^{(\prime)}$ and $K^* K$ rescattering mechanisms,
mediated by analogous $\pi$- and $K$-meson exchanges, respectively,
we follow the approach adopted in Refs.~\cite{Hsiao:2019ait,Ling:2021qzl,Yu:2021euw}
to carry out the numerical analysis. By fitting all available branching-fraction data, we obtain
$(\Lambda_\pi,\Lambda_K)=(1.1\pm 0.3,1.5\pm 0.3)$~GeV,
which are taken as the input values throughout this work. In addition, a variation
$\delta \Lambda_M=0.3$~GeV is introduced to assess the sensitivity 
of the loop calculations to the choice of cutoff.

Using these theoretical parameters, we first calculate the branching fractions of
$D\to \eta\pi^{(\prime)},K\bar K$ and $D\to a_1\eta$,
as summarized in Table~\ref{tab1}, for comparison with experimental data.
Subsequently, we calculate the rescattering branching fractions of $D\to \sigma_0 a_0$ and
$D_s^+\to f_0 a_0^+$, presented in Table~\ref{tab2}.

%
\begin{table}[t]
\caption{Branching fractions of the initial two-body $D$ meson weak decays, 
calculated and compared with experimental data. 
For $D\to (\pi\eta, \pi\eta', K\bar K)$, the uncertainties combine 
the errors of the topological amplitudes in Eq.~(\ref{TCA}), 
while for $D\to a_1 \eta$, they include the uncertainties of 
$a_{1{\rm w}}$, $a_{2{\rm w}}$, and $f_{a_1}$.}\label{tab1}
{
\scriptsize
\begin{tabular}
{|l|c|c|}
\hline
decay modes
&theoretical results
&experimental data~\cite{pdg}\\
\hline\hline
$10^2 {\cal B}(D_s^+\to\pi^+\eta,\pi^+\eta')$
&$(1.67\pm0.08,3.92\pm0.08)$
&$(1.69\pm 0.03,3.95\pm 0.08)$
\\
$10^{2}{\cal B}(D_s^+\to K^+\bar K^0)$
&$2.91\pm0.10$
&$2.95\pm 0.14$
\\
$10^3 {\cal B}(D^+\to\pi^+\eta,\pi^+\eta')$
&$(3.29\pm0.06,4.66\pm0.06)$
&$(3.77\pm 0.09,4.97\pm 0.19)$
\\
$10^4 {\cal B}(D^0\to\pi^0\eta,\pi^0\eta')$
&$(8.7\pm 3.0,13.6\pm 2.4)$
&$(6.3\pm 0.6,9.2\pm 0.1)$
\\
\hline\hline
$10^2{\cal B}(D_s^+\to a_1^+\eta)$
&$1.9\pm 0.4$
&$2.0\pm 0.9$
\\
$10^4{\cal B}(D^+\to a_1^+\eta)$
&$5.2\pm 1.3$
&---
\\
$10^6{\cal B}(D^0\to a_1^0\eta)$
&$1.6\pm 1.0$
&---
\\
\hline
\end{tabular}
}
\end{table}
%
%
\begin{table}[t]
\caption{Branching fractions of the triangle rescattering-induced $D$ meson decays,
including $D\to \pi\eta^{(\prime)}\to \sigma_0 a_0$, 
$D_s^+\to (\pi^+\eta^{(\prime)}, K^+\bar K^0)\to f_0 a_0^+$, and 
$D\to a_1 \eta\to \sigma_0 a_0$, which contribute to 
the total branching fractions ${\cal B}(D\to \sigma_0 a_0, f_0 a_0)$. 
The first quoted uncertainties arise from the input $D$-decay amplitudes 
and the strong couplings in Eq.~(\ref{couplings}), 
while the second reflect the estimated 30\% variation 
in the cutoff parameters $\Lambda_{\pi,K}$.}\label{tab2}
{
\scriptsize
\begin{tabular}
{|l|c|}
\hline
decay modes&this work\\
\hline\hline
$10^4{\cal B}(D_s^+\to\pi^+\eta\to \sigma_0 a_0^+)$
&$3.6\pm 0.3^{+0.1}_{-0.3}$
\\
$10^3{\cal B}(D_s^+\to\pi^+\eta'\to \sigma_0 a_0^+)$
&$0.9\pm 0.1^{+0.1}_{-0.2}$
\\
$10^3{\cal B}(D_s^+\to a_1^+\eta\to \sigma_0 a_0^+)$
&$5.6\pm 2.0^{+0.6}_{-0.8}$
\\
$10^2{\cal B}(D_s^+\to \sigma_0 a_0^+)$
&$1.0\pm 0.2^{+0.1}_{-0.2}$
\\
\hline\hline
$10^6{\cal B}(D_s^+\to\pi^+\eta\to f_0 a_0^+)$
&$2.6\pm 0.4^{+0.1}_{-0.9}$
\\
$10^6{\cal B}(D_s^+\to\pi^+\eta'\to f_0 a_0^+)$
&$4.0\pm 0.6^{+1.0}_{-2.6}$
\\
$10^4{\cal B}(D_s^+\to K^+\bar K^0\to f_0 a_0^+)$
&$2.3\pm0.2^{+0.2}_{-0.4}$
\\
$10^4{\cal B}(D_s^+\to f_0 a_0^+)$
&$3.4\pm 0.3^{+0.4}_{-0.9}$
\\
\hline
\end{tabular}
}
{
\scriptsize
\begin{tabular}
{|l|c|}
\hline
decay modes&this work\\
\hline\hline
$10^5{\cal B}(D^+\to\pi^+\eta\to \sigma_0 a_0^+)$
&$7.5\pm 0.6^{+0.2}_{-0.6}$
\\
$10^4{\cal B}(D^+\to\pi^+\eta'\to \sigma_0 a_0^+)$
&$1.1\pm 0.1^{+0.1}_{-0.2}$
\\
$10^4{\cal B}(D^+\to a_1^+\eta\to \sigma_0 a_0^+)$
&$4.0\pm 1.5^{+0.4}_{-0.5}$
\\
$10^3{\cal B}(D^+\to \sigma_0 a_0^+)$
&$1.1\pm 0.2^{+0.1}_{-0.2}$
\\\hline\hline
$10^6{\cal B}(D^0\to\pi^0\eta\to \sigma_0 a_0^0)$
&$10.0\pm 3.5^{+0.2}_{-0.7}$
\\
$10^5{\cal B}(D^0\to\pi^0\eta'\to \sigma_0 a_0^0)$
&$1.7\pm0.3^{+0.2}_{-0.3}$
\\
$10^6{\cal B}(D^0\to a_1^0\eta\to \sigma_0 a_0^0)$
&$1.3\pm0.9^{+0.1}_{-0.2}$
\\
$10^5{\cal B}(D^0\to \sigma_0 a_0^0)$
&$0.9\pm 0.2^{+0.2}_{-0.3}$
\\
\hline
\end{tabular}
}
\end{table}
%
\section{Discussions and Conclusion}
The $D\to SS$ decays involve  the $W$-boson emission topologies,
which, as discussed in the Introduction, are negligible.
The $W$-boson annihilation (WA) and $W$-boson exchange (WE) contributions
lead to the amplitudes~\cite{Yu:2021euw} 
${\cal M}_{\rm WA}(D_{(s)}^+\to \sigma_0 a_0^+)\simeq 
(G_F/\sqrt 2)\lambda_{d(sd)} a_{1w} f_{D_s} (m_u+m_d)
\langle \sigma_0 a_0^+|\bar u\gamma_5 d|0\rangle$ and
${\cal M}_{\rm WE}(D^0\to \sigma_0 a_0^0)\simeq 
(G_F/\sqrt 2)\lambda_{s} a_{2w} f_{D} (m_u+m_d)
\langle \sigma_0 a_0^0|\bar d\gamma_5 d|0\rangle$,
which are strongly suppressed due to $m_u + m_d \simeq 0$. 
Similarly, ${\cal M}_{\rm WA}(D_s^+\to f_0 a_0^+)$ is also negligible. Therefore,
short-distance contributions to $D\to \sigma_0 a_0$ and $D_s^+\to f_0 a_0^+$ 
are effectively zero, and any observation of these decays 
would signal significant long-distance FSI effects.

The triangle rescattering processes for $D\to \sigma_0 a_0$, 
illustrated in Fig.~\ref{fig1}$(a)$ and Fig.~\ref{fig2}$(a)$, 
are analogous to those for $D_s^+\to \rho^0 a_0^+$~\cite{Yu:2021euw}, 
with the $\rho\to \pi\pi$ coupling replaced by $\sigma_0\to \pi\pi$. Our calculations give
${\cal B}(D^+_s\to(\pi^+\eta+\pi^+\eta')\to\sigma_0 a_0^+)=(2.1\pm 0.1)\times 10^{-3}$,
where constructive interference between the $\pi\eta$ and $\pi\eta'$ amplitudes is included, 
leading to a value comparable to the measured branching fraction of 
$D_s^+\to \rho^0 a_0^+$~\cite{pdg,BESIII:2021aza}. We further obtain
${\cal B}(D^+\to(\pi^+\eta+\pi^+\eta')\to\sigma_0 a_0^+)=(2.4\pm 0.1)\times 10^{-4}$ and
${\cal B}(D^0\to(\pi^+\eta+\pi^+\eta')\to\sigma_0 a_0^+)=(1.3\pm 0.1)\times 10^{-5}$.
These results clearly demonstrate that the FSI triangle rescattering mechanism 
plays a key role in $D\to \sigma_0 a_0$ decays.

The resonant strong decay $a_1^+\to\sigma_0\pi^+$,
observed in the measured branching fraction
${\cal B}(D_s^+\to a_1^+\eta,a_1^+\to\sigma_0\pi^+,\sigma_0\to\pi^+\pi^-)$~\cite{pdg},
motivates a new rescattering contribution via $D\to a_1\eta\to\sigma_0 a_0$,
as illustrated in Fig.~\ref{fig1}(b)~and~Fig.~\ref{fig2}(b). Since the initial weak decays
$D\to a_1\eta$ have not been studied previously, we calculate them
using the amplitudes in Eq.~(\ref{amp2}),
obtaining ${\cal B}(D^+\to a_1^+\eta)=(5.2\pm 1.3)\times 10^{-4}$,
${\cal B}(D^0\to a_1^0\eta)=(1.6\pm 1.0)\times 10^{-6}$, as listed in Table~\ref{tab1}.
Notably, our result for ${\cal B}(D_s^+\to a_1^+\eta)=(1.9\pm 0.4)\times 10^{-2}$
is consistent with the experimental value of $(2.0\pm 0.9)\times 10^{-2}$,
validating the reliability of our calculations.

Based on the initial weak decays $D\to a_1\eta$,
we evaluate the corresponding rescattering contributions.
This yields ${\cal B}(D_s^+\to a_1^+\eta\to \sigma_0 a_0^+)
=(5.6\pm 2.0^{+0.6}_{-0.8})\times 10^{-3}$ and
${\cal B}(D^+\to a_1^+\eta\to \sigma_0 a_0^+)
=(4.0\pm 1.5^{+0.4}_{-0.5})\times 10^{-4}$, 
both of which are several times larger than 
the corresponding $\pi^+\eta^{(\prime)}$-rescattering contributions.
To further elucidate this enhancement, the amplitudes in Eq.~(\ref{amp_abc}), 
together with those in Eqs.~(\ref{amp1}) and (\ref{amp2}), 
can be related approximately as
\begin{eqnarray}
{\cal M}(D_s^+\to a_1\eta\to \sigma_0 a_0^+)\simeq 
{\cal R}_{\rm w}{\cal R}_g {\cal F}_{a_1}{\cal R}_{\rm Int}{\cal M}(D_s^+\to\pi\eta'\to \sigma_0 a_0^+)\,.
\end{eqnarray}
Here,  ${\cal R}_{\rm w}=
(\sqrt 2 G_F m_{a_1}a_{1{\rm w}} f_{a_1} F_1^{D_s^+\to\eta})
/(\sqrt 2 A\sin\phi+T\cos\phi)=-1.5$~GeV$^{-1}$ and
${\cal R}_g=(g_{a_1\sigma_0 \pi} g_{a_0\pi\eta})/(g_{\sigma_0\pi\pi}g_{a_0\pi\eta'})
=-1.5$~GeV$^{-1}$. The factor ${\cal F}_{a_1}=-3/2 p_0\cdot p_1=-2.4$~GeV$^2$
arises from summing over the $a_1$ polarization in the numerator of ${\cal M}_b$ 
in Eq.~(\ref{amp_abc}). In addition, the integration over 
the denominators in ${\cal M}_b$ and ${\cal M}^{\prime}_{a(\sigma_0)}$ 
yields ${\cal R}_{\rm Int}\simeq 0.4$, with the relatively larger mass 
$m_{a_1}\simeq 1.2$~GeV in ${\cal M}_b$ leading to a stronger suppression of the loop integral.
Consequently, the product ${\cal R}_{\rm w}^2{\cal R}_g^2\simeq 5$~GeV$^{-4}$ 
provides a natural enhancement factor, 
while ${\cal F}_{a_1}^2{\cal R}_{\rm Int}^2\simeq 1$~GeV$^{4}$. Taken together,
these factors account for the inequality of
${\cal B}(D_s^+\to a_1\eta\to \sigma_0 a_0^+)>{\cal B}(D_s^+\to a_1\eta^\prime\to \sigma_0 a_0^+)$.
Analogous relations can be used to interpret the inequalities 
${\cal B}(D_s^+\to a_1\eta\to \sigma_0 a_0^+)>{\cal B}(D_s^+\to a_1\eta\to \sigma_0 a_0^+)$ and
${\cal B}(D^+\to a_1\eta\to \sigma_0 a_0^+)>{\cal B}(D^+\to a_1\eta^{(\prime)}\to \sigma_0 a_0^+)$.
In contrast, ${\cal B}(D^0\to a_1^0\eta\to \sigma_0 a_0^0)
=(1.3\pm0.9^{+0.1}_{-0.2})\times 10^{-6}$ is smaller than
${\cal B}(D^0\to \pi^0\eta^{(\prime)}\to \sigma_0 a_0^0)$, 
which can be traced back to the strong suppression in the weak-interaction factor,
${\cal R}_{\rm w}^2=0.04\,(0.02)$, for the neutral channels.
By combining all contributions and including interference effects, 
we arrive at the first predictions for rescattering-induced $D\to SS$ decays
\begin{eqnarray}
{\cal B}(D_s^+\to \sigma_0 a_0^+)
&=&(1.0\pm 0.2^{+0.1}_{-0.2})\times 10^{-2}\,,
\nonumber\\
{\cal B}(D^+\to \sigma_0 a_0^+)
&=&(1.1\pm 0.2^{+0.1}_{-0.2})\times 10^{-3}\,,
\nonumber\\
{\cal B}(D^0\to \sigma_0 a_0^0)
&=&(0.9\pm 0.2^{+0.2}_{-0.3})\times 10^{-5}\,,
\end{eqnarray}
underscoring their potential observability in future experiments.

Since $m_{f_0}+m_{a_0} > m_{D^{+(0)}}$, 
the decays $D^{+(0)}\to f_0 a_0^{+(0)}$ are kinematically forbidden. 
The Cabibbo-allowed channel $D_s^+\to f_0 a_0^+$ is the only viable case; 
however, because $m_{D_s}$ lies only slightly above the $m_{f_0}+m_{a_0}$ threshold, 
the severely restricted phase space leads to strong suppression.
The relevant triangle-rescattering diagrams are shown in Fig.~\ref{fig1}(a), 
where the decay proceeds through $\pi^+\eta^{(\prime)}$ rescattering. 
In Fig.~\ref{fig1}(b), we further include the contributions from 
$D_s^+\to K^+\bar K^0 \to f_0 a_0^+$ and $D_s^+\to K^+\bar K^0 \to a_0^+ f_0$, 
mediated by $K^+$ and $\bar K^0$ exchange, respectively. 
These two identical contributions are combined, 
giving rise to the prefactor of 2 in the third amplitude of Eq.~(\ref{amp_abc}).
Summing all rescattering contributions, 
we find
\begin{eqnarray}
{\cal B}(D_s^+\to f_0 a_0^+)=(3.4\pm 0.3^{+0.4}_{-0.9})\times 10^{-4}\,.
\end{eqnarray}
Although kinematic suppression reduces this branching fraction 
to about thirty times smaller than ${\cal B}(D_s^+\to \sigma_0 a_0^+)$, 
it still reaches the level of a few $10^{-4}$. This makes 
the decay $D_s^+\to f_0 a_0^+$ a promising candidate for experimental searches.

In summary, we have presented the first systematic study of $D\to SS$ decays,
showing that short-distance contributions are negligible.
We have therefore invoked the FSI triangle-rescattering processes to enhance these decays,
such as $D\to \pi\eta^{(\prime)}\to \sigma_0 a_0$ and $D\to a_1\eta\to\sigma_0 a_0$,
where pion exchange mediates the $\pi\eta^{(\prime)}$ and $a_1\eta$ scatterings,
respectively. Our calculations yield
${\cal B}(D_s^+\to\sigma_0 a_0^+)=(1.0\pm 0.2^{+0.1}_{-0.2})\times 10^{-2}$,
${\cal B}(D^+\to\sigma_0 a_0^+)=(1.1\pm 0.2^{+0.1}_{-0.2})\times 10^{-3}$, and
${\cal B}(D^0\to\sigma_0 a_0^0)=(0.9\pm 0.2^{+0.2}_{-0.3})\times 10^{-5}$.
For the Cabibbo-allowed decay channel $D_s^+\to f_0 a_0^+$, 
the near-threshold condition $m_{D_s}\simeq m_{f_0}+m_{a_0}$
severely limits the available phase space, leading to a suppressed rate of
${\cal B}(D_s^+\to f_0 a_0^+)=(3.4\pm 0.3^{+0.4}_{-0.9})\times 10^{-4}$. 
These results demonstrate that rescattering-induced $D\to SS$ decays, 
though suppressed in some channels, remain potentially accessible in future experiments.

\section*{ACKNOWLEDGMENTS}
This work was supported in part by National Natural Science Foundation of China 
(Grants~No.~12575101 and No.~12175128).


\end{document}